# Franz-Keldysh effect in semiconductor built-in fields

22 April 2018

Yury Turkulets and Ilan Shalish*

*Department of Electrical and Computer Engineering, Ben Gurion University of the Negev, Beer Sheva, Israel*

Franz-Keldysh effect is expressed in the smearing of the absorption edge in semiconductors under high electric fields. While Franz and Keldysh considered a limited case of externally applied uniform electric field, the same effect may be caused by built-in electric fields at semiconductor surfaces and interfaces. While in the first case, the bands are bent linearly, in the latter case, they are bent parabolically. This non-linear band bending poses an additional complexity that has not been considered previously. Here, we extend the linear model to treat the case of a non-linear band bending. We then show how this model may be used to quantitatively analyze photocurrent and photovoltage spectra to determine the built-in fields, the density of surface state charge, and the doping concentration of the material. We use the model on a GaN\AlGaN heterostructure, and GaAs bulk. The results demonstrate that the same mechanism underlies the band-edge response both in photocurrent and photovoltage spectra and demonstrate the quantitative use of the model in contactless extraction of important semiconductor material parameters.

## I. Introduction

Light absorption in semiconductors commences mainly when the photon energy becomes high enough to cause band-to-band transitions. Hence, one would expect absorption to increase sharply as the photon energy becomes equal to the bandgap. However in reality, the absorption edge shows a gradual increase towards the bandgap photon energy.[1,2] It has become accepted that the main reason for this sub-band-edge response is the presence of a naturally-occurring built-in electric field at the surface.[3,4,5] In the presence of an electric field, the absorption edge is red-shifted. This effect has been classified as a case of electro-absorption or the Franz-Keldysh effect.[6,7] The original works of Franz[8] and Keldysh[9] dealt with an externally-applied electric field. An external field would usually produce a linear band-bending. The purpose of this work was to model a previously untreated case of electro-absorption in *internal built-in fields*. Unlike the linear case studied by Franz and Keldysh, built-in fields often add an additional complexity in the form of a non-linear band bending. A linear band bending may occur in semiconductors, when a layer is fully depleted, a case which is relevant in quantum size structures, e.g., quantum wells (QW) and quantum barriers. However, at semiconductor surfaces, or interfaces, of thick layers, the band-bending is typically *non-linear*. Here, we extend the electro-absorption model to the case of non-linear band bending. We model electro-absorption for both cases of internal built-in fields, with linear and parabolic band bending, and examine their manifestation in two electro-optical spectroscopies, photocurrent (also known as spectral photoconductivity), and surface photovoltage. We show that for the linear case, the model may be used for interpretation of spectral data to determine the density of charged surface states, while in the parabolic case, it may also yield the doping concentration in the layer.

We note that the Franz-Keldysh effect is known mostly for its effect *above* the bandgap photon energy (Franz-Keldysh oscillations), a phenomenon which has been used extensively in modulation spectroscopy.[10,11,12] The present work relates to its effect *below* the bandgap.

## II. Model

The main difference between the cases of linear and parabolic band-bending is that the former is formed between two equal 2D surface charges of opposite signs, while the latter is formed by a single 2D surface charge concentrated on one side, and a 3D volume charge of dopant ions on the other.

In the following treatment, we will assume that the *effective photon flux* (which is equal to the impinging photon flux multiplied by the optical transmission of the surface) is constant. Further discussion how we practically achieve this condition is given in the experimental details section.

## 1. Linear Band Bending

We consider, for example, a heterostructure made of a thin n-type AlGaN layer on top of an n-type GaN substrate. GaN and AlGaN are polar materials. Slight displacement of Ga and N atoms in c-direction, induces a dipole in each monolayer of the material. The charges forming this dipole are then canceled by charges of adjacent dipoles induced between the atoms of the adjacent layers. This nullifies the total charge inside the lattice. At the surfaces, the situation is different, as there is no further opposite charge to cancel the charge of the outmost layer. Therefore, both surfaces of the layer have equal sheet charges of opposite signs. This charge is commonly referred to as the *polarization charge*.[13,14] The AlGaN/GaN in our example is a structure commonly used for high electron mobility transistors (HEMT).[15.] The heterojunction between the AlGaN and GaN forms a triangular quantum well (QW). At equilibrium and steady state, all the surface state electrons end up in the QW due to the presence of the polarization field. The doping in the layer is typically negligible, and so the AlGaN layer is placed between two equal and opposite charges, similar to a parallel plate capacitor.

To excite an electron across the bandgap, a photon is required to have at least the energy of the bandgap. However, in the presence of a built-in electric field, photons with energies smaller than the bandgap may be able to excite an electron across the bandgap gaining the missing energy from the electric field. This process may be described as an electron excited into the forbidden gap close to the conduction band, and then tunneling sideways (or inward) into the conduction band as depicted in Fig. 1. Under a strong electric field, the Schroedinger equation solutions for the electronic states in the conduction band and in the valence band are Airy functions having their tails penetrating the forbidden gap. So the same process may be visualized as excitation taking place within the forbidden gap between the tails of the Airy functions.[16] Hence, the tunneling step shown in Fig. 1 is, in most cases, divided into two tunneling steps, one from the valence band into the forbidden gap and another from the forbidden gap into the conduction band. Since the tunneling probability is the same in both cases, we will describe, for convenience, the case shown in Fig. 1.

To calculate the tunneling probability, we will use the Wentzel–Kramers–Brillouin (WKB) approximation.[17] Figure 2 defines the energy-depth relations in the AlGaN layer of Fig. 1. Per this

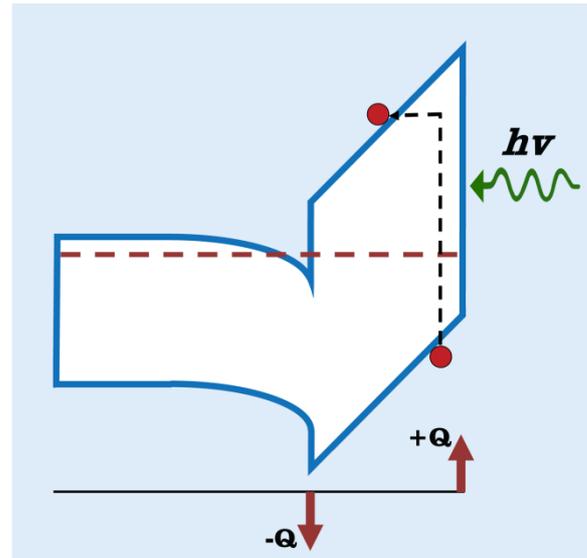

**Fig. 1** – Schematic band diagram of an AlGaN/GaN heterostructure. A photon having energy smaller than the bandgap may still generate an electron-hole pair with the assistance of the built-in field.

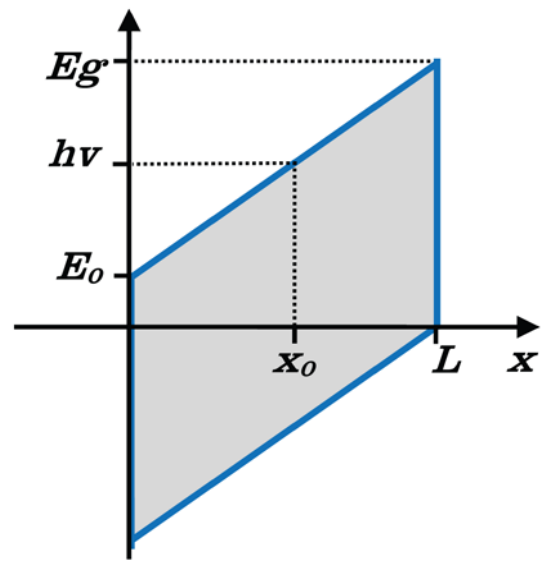

**Fig. 2** – Schematic drawing of the AlGaN layer defining the energy - depth relations. A photon energy, hv, equals the potential energy at a certain point $x_0$.

definition, the function describing the potential energy of the conduction band is

$$(1) \qquad E(x) = E_0 + \frac{E_g - E_0}{L} x$$

An electron excited from the valence band (near the surface) by a photon of energy $h\nu$ reaches a potential energy that is equal to

$$(2) \quad E(x_0) = h\nu = E_0 + \frac{Eg - E_0}{L} x_0$$

The probability of tunneling is defined as

$$(3) \quad P_{Tunnel} \approx exp\left( -2 \int_{x_0}^{L} \gamma(x) dx \right) \quad where$$
$$\gamma(x) = \frac{\sqrt{2m}}{\hbar} \left[ E(x) - E(x_0) \right]^{1/2}$$

and $m$ is the reduced effective mass, and $\hbar$ is the reduced Planck constant. Integration yields

$$(4) \quad P_{Tunnel} = exp\left[ -\left( \frac{Eg - h\nu}{\Delta E} \right)^{3/2} \right] \quad where \quad \Delta E = \left( \frac{3}{4} \frac{qFh}{\sqrt{2m}} \right)^{2/3}$$

and $F$ is the built-in electric field. This result shows that the probability to excite an electron across the bandgap does not reduce to zero at once, when the photon energy is made smaller than the bandgap, but rather decays exponentially as the tunneling probability decreases. It has been shown previously that using this expression, it is possible to model the band-edge step response in photocurrent spectra and, by fitting the photocurrent spectral data, to obtain the built-in electric field.[18] The built-in electric field is caused by the surface-state charge density, $N_T$, which is equal in our case to the charge density in the QW, $n_S$. We can, therefore, calculate this charge using the following relations: $\varepsilon F = q N_T = q n_S$.

Similarly, we will derive here a model for surface photovoltage spectra of the same structure.

## 1.1 Photocurrent Model for Linear Band-Bending

A basic photocurrent model has been presented in our previous paper.[18] It covered the case of a linear band bending. We shall now describe it briefly for the integrity of the discussion as an introduction to the more complicated case of a parabolic band bending. We will also extend it to the case of photovoltage.

When a semiconductor is exposed to monochromatic light and the photon energy is scanned from low to high energy, the conductivity typically increases as the bandgap energy is approached.[19] This band-edge photo-response should be observed as a step at the exact energy of the bandgap, where band to band electron transitions become possible. However, at the presence of a strong-enough built-in electric field, electric-field-assisted absorption enables band-to-band transitions at energies smaller than the bandgap. In the case of a uniform built-in field the photo-response takes the form of a step that is smeared to low energies. The step starts from a below-bandgap value of the current that we will denote as dark current, $I_D$, and rises gradually to reach a new level at the bandgap energy that we will denote as saturation current, $I_S$. Assuming a constant photon-flux throughout the spectrum, the actual flux reaching the sample is still reflected, in part, at the sample surface before absorption can take place. The reflectance, $R(h\nu)$, is a function of the photon energy, $h\nu$.[20] The flux available for absorption is therefore $1-R(h\nu)$ of the impinging flux. To eliminate the effect of reflectance, the acquired data has to be normalized by $1-R(h\nu)$. The normalized current step should follow the tunneling probability derived in the previous section. This probability is zero for large values of $Eg-h\nu$ (or as long as the photon energy is much smaller than the bandgap energy) and increases for decreasing values, reaching the value of $I$ at the bandgap. The current may therefore be described by the following model:

$$(5) \quad I(h\nu) = I_D + (I_S - I_D) P_{Tunnel} \quad or$$
$$\Delta I(h\nu) = \Delta I(Eg) \cdot P_{Tunnel}$$

where $\Delta I$ is the photocurrent, and $P_{Tunnel}$ is the tunneling probability, given by Eq. 4 for the case of a *linear* band bending.

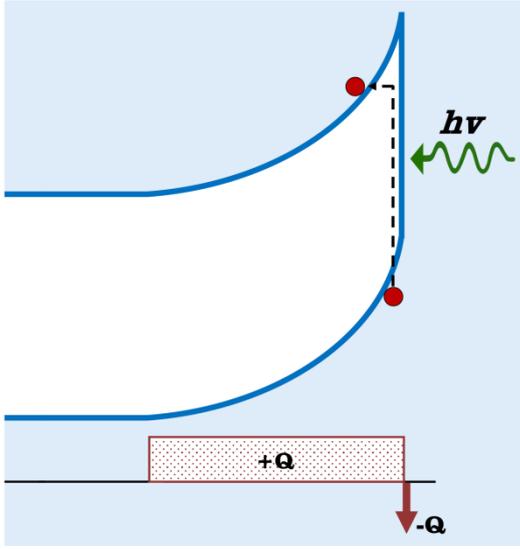

**Fig. 3** – Schematic band diagram of a depleted free surface. Here as well, a photon having energy smaller than the bandgap may still generate an electron-hole pair with the assistance of the built-in field.

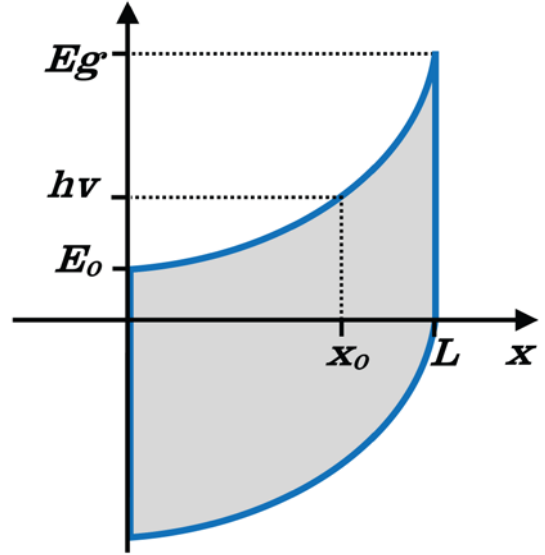

**Fig. 4** – Schematic drawing of a surface depletion layer defining the energy - depth relations. A photon energy, $h\nu$, equals the potential energy at a certain point $x_O$.

## 1.2 Photovoltage Model for Linear Band-Bending

Let us consider for example the structure shown in Fig. 1. While electrons, excited in this structure, will be swept into the well and contribute to the photocurrent, holes will be swept by the built-in fields away from the junction. We are interested in the layer with the linear band-bending, the AlGaN, wherein the holes will be swept to the surface and cancel some of the negative charge trapped at surface states. This charge separation across the layer induces a change in the electric field across the layer, resulting in a certain change to the band-bending which can be detected (e.g., using a Kelvin probe) as a photo-induced change to the surface voltage, i.e., *surface photovoltage*.[21]

The absorbed light generates electron-hole pairs, creating excess carrier density, $\Delta n(h\nu)$. On the one hand, these excess carriers (or the electrons) create the photocurrent:

(6) $\quad \Delta I(h\nu) = V \cdot q \cdot \Delta n(h\nu) \cdot \mu \frac{w}{L}$

where $V$ is the voltage between the two contacts contacting the QW, $q$ is the electron charge, $\Delta n(h\nu)$ is the photo-generated excess carriers that form excess sheet charge concentration in the 2-DEG, $\mu$ is the electron mobility in the 2-DEG, $L$ and $W$ are the length and width of the 2-DEG channel, respectively. On the other hand, the same excess carrier density creates a change in the electric field across the layer:

(7) $\quad q \cdot \Delta n(h\nu) = \varepsilon \frac{\Delta \phi_{BB}(h\nu)}{t}$

where $\Delta \Phi_{BB}(h\nu)$ is the change in the layer potential upon illumination with photons of energy, $h\nu$, and $t$ – the layer thickness. From Eqs. 6 and 7, it follows that

(8) $\quad \Delta \phi_{BB}(h\nu) = c \cdot \Delta I(h\nu)$

where $c$ is a constant. It then follows immediately from Eq. 5 that for a linear band bending,

(9) $\quad \Delta \phi_{BB}(h\nu) = \Delta \phi_{BB}(Eg) \cdot P_{Tunnel}$

## 2. Parabolic Band Bending

In many cases, such as a free depleted surface (Fig. 3), a Schottky barrier, or the wide bandgap side of a heterojunction, the built-in field is caused by an interplay between a 2D sheet charge, such as the surface-state charge, and the 3D fixed charge of dopant ions in the depletion region. Following the energy-depth relations defined in Fig. 4, the function describing the potential energy of the conduction band is

$$(10) \qquad E(x) = E_0 + (Eg - E_0)\frac{x^2}{L^2}$$

An electron excited from the valence band (near the surface) by a photon of energy $h\nu$ reaches a potential energy that is equal to

$$(11) \qquad E(x_0) = h\nu = E_0 + (Eg - E_0)\frac{x_0^2}{L^2}$$

Hence, we get

$$(12) \qquad P_{Tunnel} = \exp\left[-2\frac{\sqrt{2m}}{\hbar}\sqrt{\frac{Eg - E_0}{qL^2}}\int_{x_0}^{L}\left(x^2 - x_0^2\right)^{\frac{1}{2}}dx\right]$$

$$(13) \qquad \int_{x_0}^{L}\left(x^2 - x_0^2\right)^{\frac{1}{2}}dx = \frac{L^2}{2}\left[\sqrt{1 - \frac{x_0^2}{L^2}} - \frac{x_0^2}{L^2}\ln\left(\frac{1 + \sqrt{1 - \frac{x_0^2}{L^2}}}{\frac{x_0}{L}}\right)\right]$$

Taylor expansion of Eq. 13 with the variable $\sqrt{1 - \frac{x_0^2}{L^2}}$ yields:

$$(14) \qquad P_{Tunnel} = \exp\left[-\sum_{i=1}^{\infty}\left(\frac{Eg - h\nu}{\Delta E_i}\right)^{i + \frac{1}{2}}\right]$$

$$\Delta E_i = \left(\frac{4i^2 - 1}{2^{i+1}}\frac{\hbar q^i F^i}{\sqrt{2m}}L^{i-1}\right)^{\frac{2}{2i+1}}$$

In general, a band-edge step observed in absorption-related spectroscopies may now be fitted with a model based on this tunneling probability to obtain the maximum of the built-in electric field, $F$, and the depletion region width, $L$. This way, we can get the surface band-bending voltage $V_{BB} = F \cdot L$, the density of charged surface states $N_T = \varepsilon F/q$, and the doping concentration $N_D = \varepsilon F/(qL)$.

### 2.1 Photocurrent Model for Parabolic Band-bending

For a *parabolic* band bending, the tunneling probability is given in Eq. 14. Fitting the spectral response with a model based on this tunneling probability uses 3 fitting parameters: the bandgap, $Eg$, (usually known *a priori*), the maximum built-in field, $F$, and the depletion region width, $L$. We note that the first term in the series in Eq. 14 yields the linear case of Eq. 5. This means that if we take a rough approximation of only a single term, we will be approximating the parabolic band bending to be linear. This approximation is good only at photon energies close to the bandgap but departs from reality further away. Besides the accuracy, another advantage of using the more accurate expression is the ability to extract two additional material parameters – the doping level, and the depth of the depletion region – in addition to the field, and the surface state charge density.

### 2.2 Photovoltage Model for Parabolic Band-bending

Let us examine, for example, the case of a depleted surface of a thick semiconductor layer, which surface band bending is depicted in Fig. 3. Here, the most common scenario is of surface states that trap majority carriers, and (before illumination, i.e., in the dark) this trapped charge density, $N_{TD}$, induces an electric field that extends inward, into the layer, depleting a certain depth, $L$, which extent depends on the doping density, $N_D$, according to the relation:

$$(15) \qquad N_{TD} = N_D L$$

The following expression for the surface band bending, $\Phi_{BB}$, may be obtained from a solution of Poisson's equation

$$(16) \qquad \phi_{BB} = \frac{qN_{TD}L}{2\varepsilon} = \frac{qN_{TD}^2}{2\varepsilon N_D}$$

Upon photon absorption, a generated electron-hole pair undergoes charge separation by the built-in field in the depletion region. While the majority carriers are swept into the bulk, the minority carriers reach the surface and cancel the effect of some of the majority carriers trapped in the surface states. This way, the photo-induced excess carrier density, $\Delta n(h\nu)$, causes an equal change in the surface state charge:

$$(17) \qquad \Delta n(h\nu) = \Delta N_T(h\nu) = N_T(h\nu) - N_{TD}$$

From Eqs. 16 and 17, we get

$$(18) \qquad \frac{q}{2\varepsilon N_D}[N_{TD} + \Delta n(h\nu)]^2 = \phi_{BBD} + \Delta\phi_{BB}(h\nu) =$$
$$= \frac{q}{2\varepsilon N_D}N_{TD}^2 + \Delta\phi_{BB}(h\nu)$$

Where $\Phi_{BBD}$ is the surface potential (or surface band bending) in the dark, and $\Delta\Phi_{bb}(h\nu)$ is the photon-induced change in the surface potential at a specific photon energy, $h\nu$ (photovoltage). We get,

$$(19) \qquad \frac{2\varepsilon N_D}{q}\Delta\phi_{BB} = [N_{TD} + \Delta n]^2 - N_{TD}^2 = \Delta n(2N_{TD} + \Delta n)$$

Assuming a small perturbation ($\Delta n \ll N_{TD}$), we get $\Delta\phi_{BB} \propto \Delta n$, and this confirms the validity of Eqs. 8 and 9 for the parabolic case as well, except for the expression for $P_{Tunnel}$, for which we have to use Eq. 14. We also note that $\Delta\phi_{BB}$ here is assumed to have been normalized by $(1\text{-}R)$.

### III. Experimental Details

#### 3.1. Materials

To test the models on real spectra, we acquired photocurrent and photovoltage spectra from single layers and heterostructures of the following materials. A 11 nm thick undoped $Al_{0.3}Ga_{0.7}N$ on an undoped GaN layer on sapphire substrate was used for testing the model for linear built-in field, and Te-doped GaAs wafer was used for testing the model for parabolic band bending. The GaN substrates were grown on sapphire substrates by hydride vapor phase epitaxy (TDI inc.). The AlGaN epilayer was grown by metal-organic vapor-phase epitaxy.

#### 3.2. Experiments

Photocurrent and photovoltage spectra were acquired in the same experimental setup under identical conditions on the same samples. For spectral data acquisition, the samples were placed in a dark and shielded box at atmospheric room temperature conditions. For illumination, we used a 300 Watt Xe light source for the visible/ultra-violet range, and 250 Watt halogen lamp for the infra-red/visible range. The source light was monochromitized using a Newport Corp. double MS257 monochromator and further filtered by order-sorting long-pass filters. The beam was split and part of it was directed into an integrating sphere with photodetectors. The feedback from the photodetectors was used by a closed loop control system to maintain a constant photon flux throughout the spectral acquisition. The constant photon flux corresponded to light intensity of $3.6 \ \mu W/cm^2$ at $500 \ nm$. Photovoltage was measured using a Kelvin probe (Besoke Delta Phi Gmbh) in a dark and shielded box using the typical setup commonly used for surface photovoltage spectroscopy.[22] For photocurrent measurements, two Ohmic contacts were used. For the AlGaN/GaN structure, contacts of Ti(30nm)/Al(70nm)/Ni(30nm)/Au(100nm) were deposited by e-beam thermal deposition and annealed at 900 °C for 60 sec in nitrogen ambient. For the GaAs, contacts of Au-Ge(60nm)/Pt(20nm)/Ti(50nm)/Au(50nm) were deposited by e-beam thermal deposition and annealed at 450 °C for 30 sec in nitrogen ambient. Their Ohmic character was verified using current-voltage measurements. Resistivity and Hall effect measurements were carried out on the same samples. All measurements were carried out at room-temperature.

To acquire the photo-response of the material without the effect of the photon energy dependence of the light source, the common practice is to normalize the data to the spectral response of the lamp. This method is based on the assumption that the photo-response of the material is linear throughout the intensity range of the light source. We found that for the purpose of the experiments reported here, the intensity of the light source varied to an extent that rendered this method invalid and yielded erroneous results. Therefore, we used a home-built control system to maintain a constant photon flux throughout the spectral range of our measurements. This way, a constant number of photons per unit time and per unit area would impinge on the sample surface over the entire spectral range of the measurement. Yet, the impinging flux still had to be transmitted through the sample surface to be available for absorption. Not all the impinging photons make it through the surface. A certain part of them, $R$, is reflected from the surface. Unfortunately, the reflectance, $R$, is photon-energy dependent as well. However, its variation range is typically limited to a few percent, in which case one may safely normalize the data to the surface transmission, **I-R**.

### 3.3. Data Fitting Procedure

#### 3.3.1 Graphic method for linear band bending

This method applies to all cases of a *linear* built-in field (or cases when the band bending is parabolic but one chooses to approximate it by a linear band bending).

The photocurrent model is given by

$$(20) \qquad \Delta I(hv) = \Delta I(Eg) \cdot \exp\left[-\left(\frac{Eg - hv}{\Delta E}\right)^{3/2}\right]$$

Equation 20 may be rearranged in the following way

$$(21) \qquad y(hv) = \left[\ln\left(\frac{\Delta I(Eg)}{\Delta I(hv)}\right)\right]^{2/3} = \frac{Eg - hv}{\Delta E}$$

Presenting the data this way transforms the response into a linear curve. The linear curve intersects the photon energy axis at $hv = Eg$ and has a slope of $-1/\Delta E$. In most cases, the bandgap is known, and the fit is practically *a single parameter fit*.

The photovoltage model yields a similar equation for the linear case:

$$(22) \qquad y(hv) = \left[\ln\left(\frac{\Delta\phi_{BB}(Eg)}{\Delta\phi_{BB}(hv)}\right)\right]^{2/3} = \frac{Eg - hv}{\Delta E}$$

#### 3.2 Fitting method for parabolic band bending

This method applies to cases of a parabolic band bending. For example, to fit Eq. 9, when the tunneling probability is given in Eq. 14, we take the logarithm of both sides and get

$$(23) \qquad \ln\left[\frac{\Delta\phi_{BB}(Eg)}{\Delta\phi_{BB}(hv)}\right] = \sum_{i=1}^{\infty}\left(\frac{Eg - hv}{\Delta E_i}\right)^{i+1/2}$$

The left hand side of Eq. 23 is the fitted data, which is to be fitted with the function on the right, using the 3 fitting parameters: **Eg, F, L**. We note that in practice, we used only 2 fitting parameters, as the bandgap was known. To fit photocurrent spectra, we replace **$\Delta\phi_{BB}$** with **$\Delta I$** in Eq. 23.

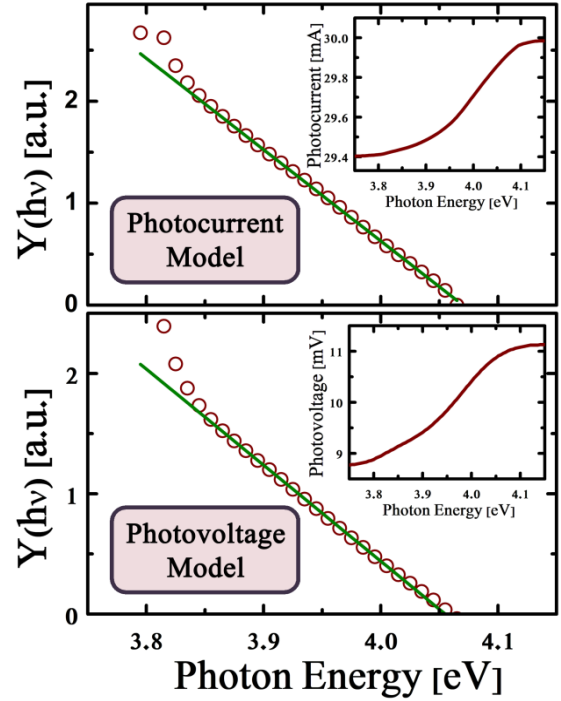

**Fig. 5 – Top –** The inset shows photocurrent step response of the AlGaN layer in the structure of Fig. 1. The step is sloped and smeared to lower energies. Using the graphic method we transform the photocurrent step into a linear curve that intersects the photon energy axis at the bandgap and which slope is -1/ΔE. **Bottom –** The inset shows the surface photovoltage band-edge step and the main curve is obtained using the graphic method. Both methods give identical results within the experimental error.

## IV. Results and Discussion

### 4.1. Linear Band-Bending

We tested the case of a linear band bending using AlGaN/GaN heterostructures for both photocurrent and photovoltage. The inset in the top panel of Fig. 5 shows the (surface-transmission-normalized) photocurrent spectrum. The main curve shows the data after applying the graphic method described in Eq. 21. The linear part of the resulting curve intersects the photon-energy axis at the transition energy (the AlGaN bandgap in this case). Eq. 4 is then used to obtain the built in electric field from the slope of the curve. Similarly, we applied the same analysis to a photovoltage spectrum obtained from the same layer, as shown in the bottom panel of Fig. 5 and obtained the same results: 1.37 ± 0.02 and 1.32 ± 0.03 MV/cm for the photocurrent and photovoltage, respectively. It is thus evident from Fig. 5 that the model describes faithfully the band edge response in both spectroscopic methods. It is also evident that there is no basic difference whether the band edge data is of photovoltage or photocurrent – the treatment is identical and the obtained parameters are identical within the experimental error.[23] As far as the band-edge response, the difference between photovoltage and photocurrent is mainly in the method of measurement. Nonetheless, the choice of the method has some important implications that need to be considered. For example, when measured with a Kelvin probe, the surface photovoltage method is *contactless* and does not require any fabrication steps. However, it is *limited to equilibrium* conditions. On the other hand, photocurrent requires at least two Ohmic contacts. However, if it is possible to apply an external field, e.g., using a gate, the photocurrent method may provide the built-in fields not only in equilibrium but also under any other external field (applied using the gate voltage).[24]

The linear portion of the curve close to the band edge provides a visual confirmation that the assumed mechanism, the Franz-Keldysh effect, is indeed the main mechanism responsible for the near band-edge smearing of the photo-response curve.

### 4.2. Parabolic Band-Bending

The case of a parabolic band bending is relevant in thick layers, e.g., in a depleted surface of a bulk sample, or in the depletion region in the wide bandgap side of a heterojunction. Here, we tested this case using GaAs. The top panel of Fig. 6 shows photocurrent and the bottom shows photovoltage spectra obtained from the same sample under identical conditions (insets). Except for the units, the spectra appear to have identical shape. Using the expression on the left hand side of Eq. 23, the spectral data is transformed to that shown as the main curves in Fig. 6 and then fitted with the series expression on the right hand side of Eq. 23. From the fit, we get the built-in field at the very surface of the sample, $F$, and the depth of the surface depletion region, $L$. The obtained maximum built-in fields are 47 ± 5.7 and 50 ± 2.9 kV/cm for photocurrent and photovoltage, respectively. The obtained depletion layer width is 20.3±4.8 and 20.5±3.6 nm for photocurrent and photovoltage, respectively. From these results, we calculated the doping using the relation $N_D = \varepsilon F / (qL)$. The obtained doping concentration was 1.65±0.44·10^17 and 1.74±0.35·10^17 cm^-3 for photocurrent and photovoltage, respectively. We also carried out Hall effect measurements on the same sample. The Hall effect carrier concentration was found to be 1.54±0.23·10^17 cm^-3. All the three methods gave values that were practically identical within the measurement error. From the value of the surface built-in field, we also calculated the density of charge in surface states using $N_T = \varepsilon F / q$. From the photocurrent data, we get 3.35±1.2·10^11 cm^-2. From the photovoltage data, we get 3.57± 0.91·10^11 cm^-2. The results are identical within the measurement error. The clear advantage of the photovoltage method is in being *contactless*.

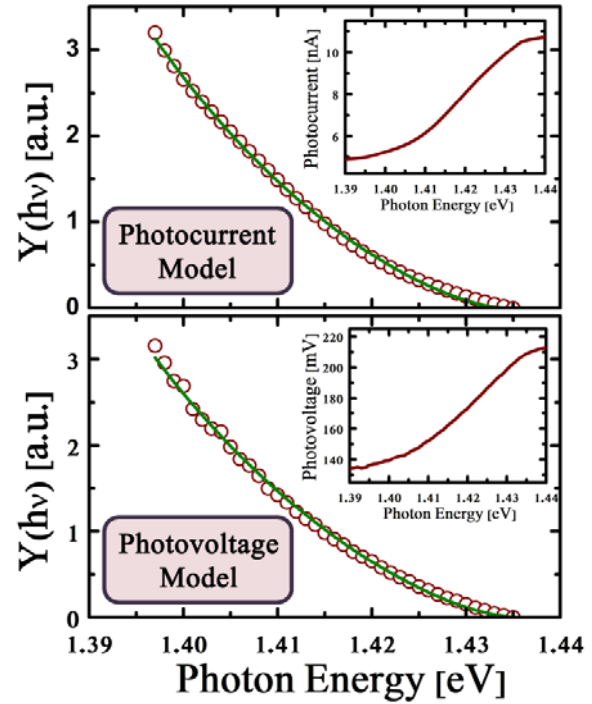

**Fig. 6** – <u>Top</u> – The inset shows photocurrent step response of the GaAs layer in the structure of Fig. 1. The step is sloped and smeared to lower energies. Using the graphic method we transform the photocurrent step into the main curve that is then used for the fit. <u>Bottom</u> – The inset shows the surface photovoltage band-edge step from the same GaAs sample and the main curve is obtained using the same graphic method. Both methods give identical results within the experimental error.

## V. Conclusion

The proposed model and the presented experimental results show the role of the surface state charge and the doping in the formation of built-in fields that give rise to the smearing of the absorption edge in spectral responses obtained using optical spectroscopies that are based on absorption. When used with surface photovoltage spectroscopy, the proposed model thus affords a contactless quantitative tool to measure carrier concentration and surface state charge density in semiconductor layers. There is currently a great difficulty in measurement of carrier concentration in nanostructures, such as nanowires, as it typically requires fabrication of single nanowire transistors. Being contactless, the combination of surface photovoltage along with an adequately modified version[25] of the proposed model may serve to assess the carrier concentration in large ensembles of nanowires at a relatively small effort.

## Acknowledgement

This work was funded by a research grant from the Israeli Ministry of Defense (MAFAT).